\documentclass[sigconf]{acmart}
\usepackage{subcaption}
\usepackage{graphicx}
\usepackage{booktabs}

\AtBeginDocument{%
  }

\setcopyright{rightsretained}
\copyrightyear{2025}
\acmYear{2025}
\acmDOI{TBD}

\acmConference[RecSys '25]{TBD}
\acmBooktitle{RecSys ’25.}

\graphicspath{ {./images/} }
\usepackage{enumitem}
\usepackage{float}

\begin{document}

\title{Addressing Cold Start For next-article Recommendation}

\author{Omar Elgohary}
\affiliation{%
\orcid{0009-0004-2778-5423}
  \institution{University of Minnesota}
  \city{Minneapolis}
  \country{USA}}
\email{elgoh003@umn.edu}

\author{Trenton Marple}
\affiliation{%
  \institution{University of Minnesota}
  \city{Minneapolis}
  \country{USA}}
\email{marpl024@umn.edu}

\author{Nathan Jorgenson}
\affiliation{%
  \institution{University of Minnesota}
  \city{Minneapolis}
  \country{USA}}
\email{jorge741@umn.edu}

\begin{abstract}
This replication study modifies ALMM, the Adaptive Linear Mapping Model constructed for the next song recommendation, to the news recommendation problem on the MIND dataset. The original version of ALMM computes latent representations for users, last-time items, and current items in a tensor factorization structure and learns a linear mapping from content features to latent item vectors. Our replication aims to improve recommendation performance in cold-start scenarios by restructuring this model to sequential news click behavior, viewing consecutively read articles as (last news, next news) tuples. Instead of the original audio features, we apply BERT and a TF-IDF (Term Frequency-Inverse Document Frequency) to news titles and abstracts to extract token contextualized representations and align them with triplet-based user reading patterns. We also propose a reproducibly thorough pre-processing pipeline combining news filtering and feature integrity validation. Our implementation of ALMM with TF-IDF shows relatively improved recommendation accuracy and robustness over Forbes and Oord baseline models in the cold-start scenario. We demonstrate that ALMM in a minimally modified state is not suitable for next news recommendation.

\end{abstract}



\maketitle
\section{Introduction}
Automated content recommendation is a sophisticated attribute of advanced information systems that assists users in identifying articles they are interested in, which are dynamically updated. Most traditional collaborative filtering approaches exploit user preferences through social connections but suffer from cold start limitations, especially when content that is newly available or infrequently engaged is proposed, such as in \cite{hu2008collaborative, koren2009matrix}. Content based approaches partially mitigate the problem considering article attributes, however, they do not capture fully user behavior's sequential and contextual aspects as suggested in \cite{wang2014improving}.

To solve the problem, we replicate and adapt the ALMM developed for next-song recommendation \cite{chou2016almm} to news recommendation. ALMM performs tensor factorization while integrating cold-start capable dynamically adjusted mapping of content-feature representative items with their latent descriptions. In the original work, song transition was sequentially modeled with audio features. In our case, we change the data model to sequential news reading data from MIND dataset \cite{wu2020mind}, where users' impressions are regarded as a time-ordered sequence of interacting with articles.

Another aspect of our approach is using contextualized language representations from BERT~\cite{devlin2019bert} 
 and TF-IDF in place of audio features. These embeddings facilitate more expressive modeling of article content, which enhances generalization for unseen news. We improve the triplet generation scheme using user sequential clicks (last\_news, next\_news), estimating missing article publication dates, and ensuring cross-component data consistency.

Our findings demonstrate that TF-IDF-augmented ALMM yields better recommendation quality than Forbes and Oord specifically in cold-start performance. Though, general scores for all algorithms were undesirable.

\section{Dataset Construction}
We build our next-article recommendation dataset using the Microsoft News Dataset (MIND)~\cite{wu2020mind}, a large-scale benchmark for personalized news recommendation. MIND contains rich textual metadata (title, abstract, entities) and user behavior logs in the form of impression sessions.

\subsection{Preprocessing and Filtering}

We first process the \texttt{news.tsv} and \texttt{behaviors.tsv} files. Articles containing duplicate or missing IDs have been removed, and all URLs are deleted because of being frequently dead. Each article is uniquely identified by its news ID, which corresponds to the stored title and abstract. In \texttt{behaviors.tsv}, we obtain click streams for every user and reorganize them into time-aligned reading sessions.

In order to maintain transition quality, we only keep consecutive pairs of articles where the time interval between the two clicks does not exceed half an hour, as in prior work~\cite{chou2016almm}. Additionally, quite short sessions and self-transitions are eliminated. This creates a transition-inclusion filtered dataset in which each user-article interaction is described as a triplet: user $u$, last-clicked article $i$, and next-clicked article $j$.

\subsection{Triplet Dataset Generation}

We create training data in the form of triplets \( (u , i , j) \) such that \( u \) corresponds to the User ID, \( i \) corresponds to the previous article read which is an input to the model, and \( j \) corresponds to the next article that was clicked on. These are extracted from filtered click sequences using a sliding window and within certain temporal bounds, self-transitions are removed.  

Instead of applying the same score of confidence to all transitions, we assign a score based on the specific transition for article $i$ and $j$. Each article is first assigned the base confidence of 1.0. Then for every user specific move of going from i to j in document is seen once, this raises the document specific confidence by 0.1. We preferred this new confidence configuration as it focuses on repeated article rotations happening with all users unlike single user based one in the original ALMM paper. This is mostly due to a domain difference: Music listeners often play the same exact song multiple times while news users usually read an article once. 

We devise train/test splits so that test triplets have at least one article which has not been trained on, enabling evaluation in cold start scenarios.

\subsection{Feature Extraction}

\textbf{(1) TF-IDF Embeddings:} We first tokenize and clean each article’s title and abstract, and apply standard TF-IDF vectorization over the texts. The resulting sparse vectors capture surface-level term importance and are used as input to content-based baselines for comparison.

\textbf{(2) BERT Embeddings:} For richer semantic representation, we use a pre-trained BERT encoder. Each article’s title and abstract are concatenated, tokenized, and passed through BERT. We extract the final hidden states and compute a mean-pooled vector over the token embeddings. This yields a dense contextualized vector $\mathbf{A}_i \in \mathbb{R}^{m}$ for article $i$, which serves as input for the ALMM and content-aware matrix factorization models.

\section{Next-Article Recommendation}

\subsection{Formalization}

The goal of next-article recommendation is to select and recommend a relevant news article to the user based on their latest click. Let $\mathcal{U} = \{u_1, u_2, ..., u_{|\mathcal{U}|} \}$ be the user set and $\mathcal{N} = \{n_1, n_2, ..., n_{|\mathcal{N}|} \}$ be the set of news articles. For each user $u$, there is a reading sequence $R^u = (r^u_1, r^u_2, ..., r^u_T)$, where each $r^u_t \in \mathcal{N}$ is a news article that was clicked at time $t$.  

To model sequential patterns, for each user reading log, we form transition pairs $(r^u_{t-1}, r^u_t)$ and filter out self-transitions, taking into account that the time interval between two consecutive clicks is no greater than thirty minutes. This creates a transition tensor $P \in \mathbb{R} ^{|\mathcal{U}| \times |\mathcal{N}| \times|\mathcal{N}|}$ , where $P^u_{i,j}$ shows the count of observed transitions, for user $u$, from article $n_i$ to $n_j$. We build a user-centric transition tensor that encodes pairwise article transitions based on user's click streams. Our method, in contrast to the ALMM formulation~\cite{chou2016almm} which captures transitions on grouped songs in sequential playlists, focuses on individual article click timestamps and strict order.

A transition $(i, j)$ is counted only if it satisfies the following criteria:
\begin{itemize}
    \item Article $i$ was clicked immediately before article $j$,
    \item The time interval between the two clicks is less than or equal to 30 minutes,
    \item The articles are distinct: $i \ne j$.
\end{itemize}

The formal definition of the transition tensor is:

\[
P^u_{i,j} = \left| \left\{ (r^u_{t-1}, r^u_t) \,\middle|\, r^u_{t-1} = i,\, r^u_t = j,\, \Delta t \leq 30\,\text{minutes},\, i \ne j \right\} \right|
\]

Here, $r^u_t$ denotes the article clicked by user $u$ at timestamp $t$, and $\Delta t$ is the time between the two consecutive clicks.

This formulation differs from the original ALMM design, which constructs transitions from adjacent sets of items by counting all pairwise combinations between them:

\[
P^u_{i,j} = \sum_{n=1}^{T-1} \left| \left\{ (L^u_n, L^u_{n+1}) \,\middle|\, i \in L^u_n \land j \in L^u_{n+1} \right\} \right|,
\]
where $L^u_n$ and $L^u_{n+1}$ are sets of songs played by user $u$ in adjacent sessions.

 Our formulation places focus on the order of actions and time of executing them while reading news articles. It ensures that transitions are actually based on paths that users take for navigation, which improves the trustworthiness of the tensor in representing user intent and enhances suitability for domains such as news where sessions are sequential and time-sensitive.

\subsection{Content-based Next-Article Recommendation}
\subsubsection{Baseline Content-Based Approaches}

\textbf{Forbes~\cite{forbes2011content}:} This method learns latent article vectors as linear transformations of content features during matrix factorization. Gradients are computed for all parameters and updated via stochastic gradient descent.

\textbf{Oord~\cite{oord2013deep}:} This method decouples latent factor learning and feature mapping. First, the user and article latent vectors are trained using standard pairwise factorization without content. Then, content-to-latent mappings are learned by minimizing the reconstruction error. At inference time, predictions use mapped features.

\begin{table}[h]
\centering
\caption{Statistics of Warm-Start and Cold-Start Data Splits}
\label{tab:data_splits}
\begin{tabular}{lccc}
\toprule
\textbf{Data sets} & \textbf{\#users} & \textbf{\#items} & \textbf{\#entries} \\
\midrule
Train (WS) & 8,485  & 21,517 & 238,038 \\
Test (WS)  & 59,047 & 11,809 & 59,047  \\
Train (CS) & 19,400 & 18,899 & 218,635 \\
Test (CS)  & 2,989  & 1,140  & 5,171   \\
\bottomrule
\end{tabular}
\end{table}

\subsubsection{Adaptive Linear Mapping Model (ALMM)}
We implement the Adaptive Linear Mapping Model (ALMM)~\cite{chou2016almm} to jointly learn user representations with spatial embeddings of the content for next-article recommendation. The model concurrently learns the user embeddings $\mathbf{U}_u$, the latent vector for the last clicked article $(\mathbf{X}_i)$, the next clicked article $\mathbf{Y}_j$, and the mapping matrices $\Psi_X, \Psi_Y$ which project article features into the latent space.

ALMM, at each iteration, performs the following steps in their defined set in each iteration:

\begin{itemize}[noitemsep]
    \item Compute new latent vectors using alternating least squares,
    \item Update the mapping matrices using ridge regression to learn the content features over the latent space,
    \item Refresh the representations of the articles with the newly mapped features.
\end{itemize}

The prediction score for a triplet $(u, i, j)$ is computed as:

\[
\hat{C}^u_{i,j} = \mathbf{U}_u^\top \mathbf{X}_i +  \mathbf{U}_u^\top \mathbf{Y}_j +  \mathbf{X}_i^\top \mathbf{Y}_j 
\]

 Each alternating step guarantees that content features are well constrained to factors learned during modeling. To extend to cold-start articles which were not trained on, we execute the learned $\Psi_Y$ at inference to yield $\mathbf{Y}_j$ directly from the content and so, generalize to new articles.

Each approach: Forbes, Oord and ALMM achieve fast inference by applying learned mappings to the vectors of article content assigned to them. We conduct the experiments using both TF-IDF and BERT embeddings as inputs to $\Psi$, so we can measure its efficiency and examines the impact of contextual features on recommendation accuracy, and a set of evaluation metrics.

\section{Evaluation Metrics}
We compare the proposed ALMM model with the following baselines:

\begin{itemize}[noitemsep,topsep=2pt]
    \item \textbf{Forbes}~\cite{forbes2011content}: A linear content-based model that jointly learns content mappings and latent vectors.
    \item \textbf{Oord}~\cite{oord2013deep}: A two-stage model where latent vectors are first learned independently, followed by post-hoc mapping from article features.
\end{itemize}

All models use TF-IDF and BERT features and share the same latent dimension and regularization settings for fair comparison.

\subsection{Evaluation Metrics}

We adopt the following ranking and qualitative metrics:

\begin{itemize}[noitemsep,topsep=2pt]
    \item \textbf{MAP@K}: Mean Average Precision at $K$, evaluating ranking quality.
    \item \textbf{Recall@K}: Measures coverage of actual next items among top-$K$ predictions.
    \item \textbf{Novelty}: Encourages recommending less frequently seen articles, computed as the inverse log popularity.
    \item \textbf{Diversity}: Assesses intra-list dissimilarity using cosine distances of TF-IDF vectors.
\end{itemize}

\section{Results}

\begin{table}
\centering
\caption{Standard and Cold-Start Evaluation Results (MAP@10, Recall@10, MAP@20, Recall@20)}
\label{tab:results}
\begin{tabular}{lcccc}
\toprule
\textbf{Model} & \textbf{MAP@10} & \textbf{Recall@10} & \textbf{Map@20} & \textbf{Recall@20} \\
\midrule
\multicolumn{5}{c}{\emph{Standard Evaluation}} \\
ALMM      & 0.0001 & 0.0003 & 0.0001 & 0.0010 \\
Forbes    & 0.0010 & 0.0021 & 0.0010 & 0.0030 \\
Oord      & 0.0007 & 0.0023 & 0.0008 & 0.0031 \\
\midrule
\multicolumn{5}{c}{\emph{Cold-Start Evaluation}} \\
ALMM      & 0.0005 & 0.0021 & 0.0006 & 0.0030 \\
Forbes    & 0.0001 & 0.0004 & 0.0002 & 0.0015 \\
Oord      & 0.0001 & 0.0015 & 0.0002 & 0.0026 \\
\bottomrule
\end{tabular}
\end{table}

We evaluate the performance of the model on both the standard and cold-start conditions using the following metrics: MAP, Recall, Novelty, and Diversity. A comparison of ALMM with Forbes and Oord models, using the same data splits, is provided in Table~\ref{tab:results}.

\subsection{Standard Recommendation Setting}

Forbes ranks the highest among the other models in the standard setting achieving MAP@10 = 0.0010 and Recall@10 = 0.0021. This is closely followed by Oord with MAP@10 = 0.0007, Recall@10 = 0.0023. ALMM performs poorly with MAP@10 = 0.0001 and Recall@10 = 0.0003, suggesting it is poorly calibrated for in-distribution predictions. These results demonstrate the edge Forbes and Oord have for common patterns and frequent transitions for their training data. 

When examining content novelty and diversity, ALMM performs better in larger recommending sets although it is not able to reach the levels set by Forbes. Forbes demonstrates the highest novelty (reaching above 3.5) and diversity followed by ALMM showing much lower novelty (below 0.5 at 500) and moderate diversity. This tradeoff demonstrates ALMM's joint learned content mappings and latent factors under standard settings lack the ability to effectively prioritise rare or distinct content that are needed.

\subsection{Cold-Start Setting}

In the cold-start setting where at least one article in each test triplet was excluded from training, ALMM demonstrates stronger generalization performance. Forbes and Oord are both surpassed in MAP@10 (0.0005) and Recall@10 (0.0021) by ALMM, which also slightly bests them in Recall@20 as well. Forbes and Oord suffer significantly more, with Forbes dropping to Recall@10 = 0.0004 and Oord to 0.0015. These results support the hypothesis that ALMM’s mapping matrices ($\Psi_X$, $\Psi_Y$) heuristically outperform other content features when predicting out-of-scope items for content features. 

ALMM, however, still trails behind Forbes in novelty and diversity. While ALMM’s recommendations tend to cluster too much around a small section of the item space fore most categories. That said, ALMM’s cold-start endurance showcases its practicality for fast-paced contexts like online news, where fresh articles are commonly published.

We also tried swapping TF-IDIF vectors with BERT-based article embeddings. Strikingly, all models performed worse with BERT features from both qualitative and ranking measures. This decline in performance defaults can be traced back to the soaring noise inflicted by BERT's high dimensionality which probably contradicts the latent factor models’ reliance on noise-free, linear architectures. Thus, while BERT provides expressive representations, its potential may not be realized until with this model.

\begin{figure*}[htbp]
    \centering

    \begin{subfigure}[t]{0.45\textwidth}
        \centering
        \includegraphics[height=4.5cm]{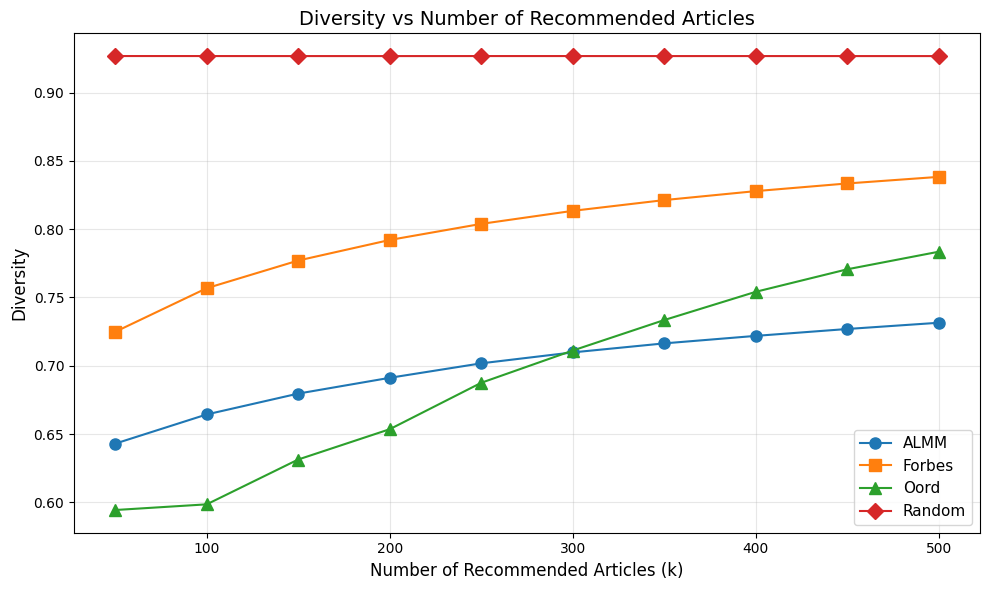}
        \caption{Diversity}
        \label{fig:diversity}
    \end{subfigure}
    \hfill
    \begin{subfigure}[t]{0.45\textwidth}
        \centering
        \includegraphics[height=4.5cm]{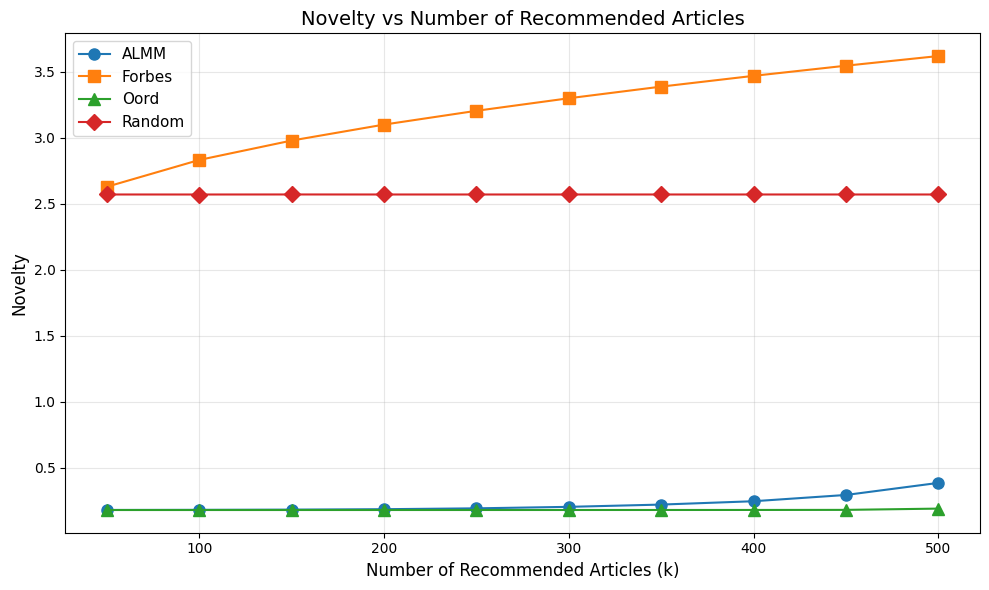}
        \caption{Novelty}
        \label{fig:novelty}
    \end{subfigure}

    \begin{subfigure}[t]{0.45\textwidth}
        \centering
        \includegraphics[height=4.5cm]{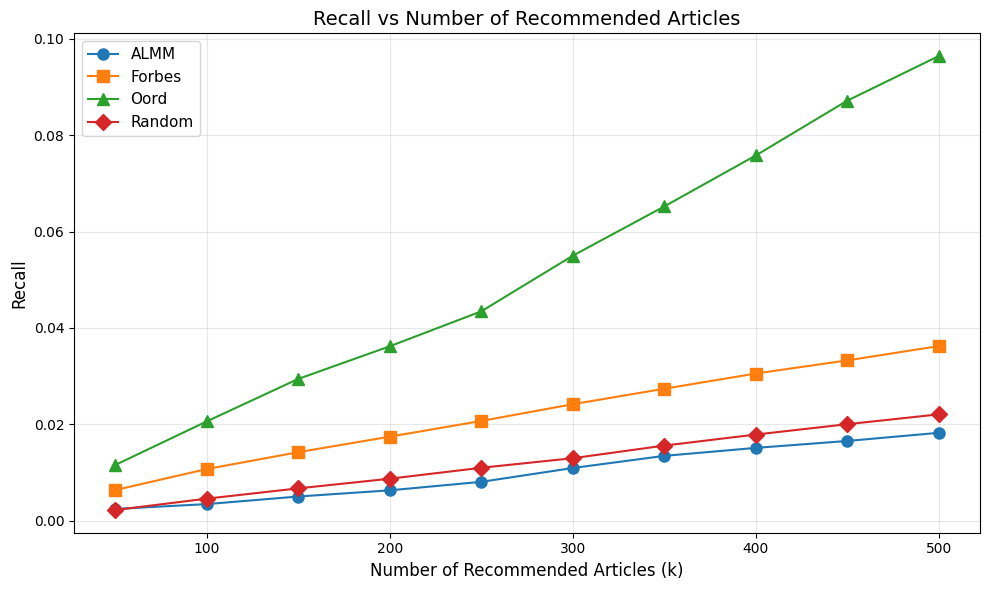}
        \caption{Recall}
        \label{fig:popularity}
    \end{subfigure}

    \caption{Performance of models across different recommendation metrics as the number of recommended articles increases.}
    \label{fig:all_metrics}
\end{figure*}

\section{Discussions}

The results reveal important trade-offs among ALMM, Forbes, and Oord across both standard and cold-start recommendation scenarios. Although ALMM demonstrates relative improvements over baseline algorithms in cold-start scenarios, the overall performance across all models is remarkably weak. Absolute values for MAP@10 and Recall@10 are extremely low and well below typical thresholds for acceptable recommendation performance which suggests that these collaborative filtering methods may be ill-suited for next-article prediction in news environments.

In the standard setting, Forbes achieves the highest MAP@10 (0.0010) and Recall@10 (0.0021), followed by Oord. ALMM lags behind in both metrics (e.g., MAP@10 = 0.0001), indicating limited capacity to model short-term user preferences or popular article transitions. 

Despite these shortcomings, ALMM demonstrates clear advantages in the cold-start scenario, where articles in the test set are unseen during training. In this setting, ALMM outperforms both Forbes and Oord across MAP@10, Recall@10, and MAP@20. These improvements, though numerically small, are meaningful in relative terms and highlight ALMM's structural strength in handling unseen content through its integrated content mapping approach. This performance stability under domain shift suggests that ALMM’s adaptive learning mechanism may transfer well to other cold-start problems beyond music, as originally intended, and into recommendation tasks in fields such as news, education, or e-commerce when built in a heavily refined and overhauled manner.

Qualitative metrics tell a more nuanced story. Forbes excels in novelty and diversity, ranking and surfacing more unique content as $k$ increases due to its stronger content regularization. ALMM trails in novelty but steadily improves in diversity, suggesting better overall content coverage but weaker ability to prioritize rare items. Oord ranks in the middle, performing well in recall but poorly in both diversity and novelty, indicating its tendency to recommend frequent, similar items.

Additionally, we explored using BERT-based embeddings in place of TF-IDF. Surprisingly, all models performed worse when BERT features were used especially ALMM. We attribute this to the high-dimensional which likely introduced noise and misalignment when passed through shallow latent factor architectures like those used here. This outcome further emphasizes that naively substituting more powerful embeddings without structural alignment to the model's architecture can be counterproductive.

In sum, while none of the tested models are strong candidates for real-world news recommendation based on their absolute performance, the relative strength of ALMM under cold-start conditions offers a promising insight. It demonstrates the value of learning content mappings during factorization, particularly in domains characterized by sparse or rapidly evolving content. Future work should investigate how to build on this strength with more sophisticated architectures that are capable of handling complex embeddings and capturing short-term user intent in high-turnover domains like news.

\subsection{Limitations}
Throughout our replication, one limitation that we discovered was due to the nature of the online news domain. Unlike in the music domain, where users replay the same song multiple times, users in the MIND data set only read the same article once. Because of this, we were not able to calculate accurate confidence scores on a per-user basis as in the original paper. We also didn't have access to the publishing date of the articles that were in the dataset, which made us unable to properly calculate the freshness metric. We tried web scraping using the article titles to find the dates, but too many of the articles no longer exist. Also, due to time constraints, we were only able to try three ways to calculate confidence scores for our triplets. With further trial and error, we could have potentially improved results. 

We also chose not to include our popularity metric because we felt that our implementation was not as accurate as we would have liked.

\section{Conclusion}
Testing ALMM and the two baseline algorithms in the news domain with the MIND data set showed poor metric performance in every category. Our findings did show that \textbf{relative} to the baseline algorithms, ALMM was able to significantly increase cold-start performance from the baseline. Despite this, our test of ALMM has shown that it will not transfer directly from the music domain to online news recommendation. As mentioned in the discussions section, we believe that with a significant overhaul, the adaptive learning mechanism in ALMM could be useful to domains outside of music next-song recommendation. We also believe that the performance of ALMM was hindered in our implementation due to a lack of content features to extract latent vectors from the dataset we utilized. We also believe that there could be a better way to extract the latent vectors for use with ALMM. We believe that the failure of BERT to generate useful results cohesively with ALMM was due to not being designed for the high-dimensional latent features generated by BERT. ALMM with TF-IDF was able to provide us with usable results, but they were still subpar. Overall, ALMM is not fit for use within the news recommendation domain, but could be modified to increase performance.

\bibliographystyle{ACM-Reference-Format}
\bibliography{citations}
\newpage
\onecolumn

\end{document}